\documentclass[review,times,authoryear]{elsarticle}
\usepackage{soul} % Hightlight text

\usepackage{xcolor}
\usepackage{framed}
\colorlet{shadecolor}{yellow}
\newenvironment{New}{}{}
\usepackage{amsmath,amssymb,amsfonts,amsthm}
\usepackage{algorithmic}
\usepackage{graphicx}
\usepackage{textcomp}
\usepackage{xcolor}

 \usepackage{siunitx}
\newcommand{\BGT}{\texttt{BondGraphTools }}

\definecolor{codebg}{HTML}{F6F6F6}
\definecolor{codeframe}{HTML}{CCCCCC}
\usepackage{listings}
\usepackage[hidelinks]{hyperref}
\usepackage{url,doi}
\newcommand{\BG}[1]{\text{\sffamily\textbf{#1}}}

\newcommand{\C}{\BG{C }}
\newcommand{\Ce}{\BG{Ce }}

\newcommand{\one}{\BG{1 }}
\newcommand{\zero}{\BG{0 }}

\renewcommand{\Re}{\BG{Re }}

\newcommand{\BGL}[2]{$\BG{#1}$:$\mathbf{#2}$} %Generic

\newcommand{\BC}[1]{\BGL{C}{#1}}
\newcommand{\BCe}[1]{\BGL{Ce}{#1}}

\newcommand{\BR}[1]{\BGL{R}{#1}}

\newcommand{\BRe}[1]{\BGL{Re}{#1}}

\usepackage{amsmath,amssymb,amscd,amstext,mathtools,extarrows,centernot}
\numberwithin{equation}{section}

\newcommand{\lb}{\left (}
\newcommand{\rb}{\right )}
\newcommand{\Phif}{{\Phi^f}}
\newcommand{\Phir}{{\Phi^r}}

\usepackage{listings}
\usepackage{chemformula}

\newcommand{\Std}{\ominus}
\newcommand{\std}{\oslash}

\usepackage{graphicx,subfigure,fancybox,color}

\newcommand{\Fig}[2]{
  \includegraphics[width=#2\columnwidth]{Figs/#1.pdf}
   \label{subfig:#1}
}

\newcommand{\SubFig}[3]{
  \subfigure[#2]{
    \includegraphics[width=#3\columnwidth]{Figs/#1.pdf}
    \label{subfig:#1}
  }
}

\begin{document}
\begin{frontmatter}
\title{Physically-Plausible Modelling of Biomolecular Systems: \\A Simplified, Energy-Based Model of the Mitochondrial Electron Transport Chain}
%%\author{Peter Gawthrop \IEEEmembership{Senior Member, IEEE}}
%%\usepackage[noblocks]{authblk}
\author[1,2]{Peter J. Gawthrop\footnote{Corresponding
    author. \textbf{peter.gawthrop@unimelb.edu.au}}}
\author[1,3]{Peter Cudmore}
\author[1,2,3]{Edmund J. Crampin}
\address[1]{Systems Biology Laboratory,
    Department of Biomedical Engineering,
    Melbourne School of Engineering,
    University of Melbourne,
    Victoria 3010, Australia}
\address[2]{Systems Biology Laboratory,
    School of Mathematics and Statistics,
    University of Melbourne, 
    Victoria 3010, Australia}
 \address[3]{ARC Centre of Excellence in Convergent Bio-Nano Science and Technology, 
    School of Chemical and Biomedical Engineering,     
    Melbourne School of Engineering,
    University of Melbourne, 
    Victoria 3010, Australia}

  \begin{abstract}
    Advances in systems biology and whole-cell modelling demand
    increasingly comprehensive mathematical models of cellular
    biochemistry. Such models require the development of simplified
    representations of specific processes which capture essential biophysical
    features but without unnecessarily complexity. Recently there has
    been renewed interest in thermodynamically-based modelling of
    cellular processes. Here we present an approach to developing of
    simplified yet thermodynamically consistent (hence physically
    plausible) models which can readily be incorporated into large
    scale biochemical descriptions but which do not require full
    mechanistic detail of the underlying processes. We illustrate the
    approach through development of a simplified, physically plausible
    model of the mitochondrial electron transport chain and show that
    the simplified model behaves like the full system.
\end{abstract}

% \begin{keyword}
%   Bond graph;
%   Biochemistry;
%   Chemical reaction network;
%   Biomedical engineering;
%   Systems biology
% \end{keyword}
\end{frontmatter}

%\maketitle

% \newpage
% \tableofcontents
%%\newpage
% %%\doublespacing

\section{Introduction}
\label{sec:introduction}
%% Models
In mathematical biology, and more widely, the relative merits of simple `toy' models, which represent some key aspects of the system but not full mechanistic detail, and comprehensive mechanistically detailed representations have long been debated. Simple `toy' models allow rigorous mathematical analysis and are generally easy to simulate, but are difficult to relate to the full system and measurements thereof. Full mechanistically detailed models on the other hand provide a straight-forward mapping to the real system, but are challenging to parameterize and analyse, and may require significant computational overhead to simulate. 

%% Simple models
Simple models of complex biochemical processes can elucidate basic
behaviour and biologically significant trade-offs
\citep{ScoKluMat14,WeiOyaDan15} and as such can be used as an aid to
synthetic biology \citep{DarKimJim18}. 
%% Components of large multiscale models
Furthermore, models of individual processes may be used as part of a model of an
overall system as, for example, in the Physiome Project~\citep{crampin_multi-scale_2004,Hun16}, or in whole-cell modelling~\cite{KarSanMac12,MacRugCov14}; this requires models to be modular and reusable
\citep{NeaCooSmi14,NicAtaBon16}. 

Recently there has been renewed interest in thermodynamically-based mechanistic modelling of cellular processes \citep{MasCov19,PanGawTra19,GawSieKam17,KliLieWie16,BeaQia10}. A modular approach to energy-based modelling has been developed in the context of biomolecular systems \citep{GawCurCra15,GawCra16}. This raises the question as to whether it is possible to develop energy-based models that are nevertheless simple. 

%Mathematical models are an essential adjunct to scientific discovery
%and application \citep{Gel16}.
%
%Methods for creating mathematical models in the context of
%biomolecular systems are well established and relevant textbooks
%include those of \citet{Gol96,Sav09,KeeSne09,BeaQia10,IglIng10,Bea12,Ing13,Voi13,Sau14,KliLieWie16}.

%% Bond Graph  approach
Like engineering systems, living systems are subject to the laws of
physics in general and the laws of thermodynamics in particular.
This fact gives the opportunity of applying engineering approaches to
the modelling, analysis and understanding of living systems.
The bond graph method of \citet{Pay61} is one such well-established
engineering approach~\citep{Cel91,GawSmi96,GawBev07,Bor10,KarMarRos12}
which has been extended to include biomolecular
systems~\citep{OstPerKat71,OstPerKat73,GawCra14,GawCra17,GawSieKam17,GawCra18a,GawCra18,PanGawTra18,PanGawTra19}.

%% Physically plausible
When developing simplified models of biomolecular systems where energy transduction is important, it is essential that models be \emph{physically-plausible}.
A \emph{physically-plausible} model of a physical system has two attributes:
it is itself a model of a physical system (i.e. it does not contravene the laws of physics);
and it shares key behaviours with the actual physical system \citep{Gaw03a}.
%Typically, the physically-plausible model will be much simpler than a full model of the physical system.
Such an approach will, however, only be of use if there are complex physical systems which can indeed be represented by a simpler physical model. 
This paper shows that this is indeed the case. 
In particular we demonstrate that it is possible to develop a simplified model of the mitochondrial electron transport chain that is thermodynamically consistent, but which doesn't represent full mechanistic detail, and show that it behaves like the full model. 

Mitochondria make use of reduction-oxidation (redox) reactions in which the transfer of electrons is used to provide the power driving many living systems. 
As discovered by \citet{Mit61,Mit76,Mit93,Mit11}, the key feature of
mitochondria is the \emph{chemiosmotic} energy transduction whereby a
chain of redox reactions pumps protons across the mitochondrial inner
membrane to generate an electrochemical gradient known as the \emph{proton-motive force} (PMF). The PMF is
then used to power the synthesis of ATP -- the universal fuel of
living systems.
Due to this central role in living systems, mathematical modelling of the key components of mitochondria is thus an important challenge to systems biology.
Because mitochondria transduce energy, an energy-based modelling
method is desirable, and Beard and colleagues have developed the most comprehensive such models to date \citep{Bea05,WuYanVin07,BeaQia10,Bea12,BazBeaVin16}. 
A bond graph model of mitochondrial oxidative phosphorylation has been
given by \citet{Gaw17a}. This model is based on modelling the redox
reactions associated with complexes \ch{CI}, \ch{CIII} and \ch{CIV} of
the mitochondrial electron transport chain.

%The advantages of having a simpler model are:
%%
%it is easier to understand a simple model than a complex model;
%%
%the computational and numerical aspects of simulation are eased;
%%
%it may be possible to develop explicit formulae to avoid simulation;
%%
% and fitting to experimental data involves fewer parameters.

%The advantages of a physical model are:
%%
%the parameters of a physical model have a clearer interpretation than
%those of a purely empirical model;
%%
%the behaviour of the model can be understood in physical terms;
%%
%models can be combined via energy-transducing ports and
%%
%thus non-physical behaviours incompatible with thermodynamic principles
%are eliminated.

%As bond graph  models correspond to physical systems, they proved a
%natural formulation of physically-plausible models.

Below 
%~\ref{sec:bond-graph-approach} 
we briefly outline the bond graph approach to
modelling energy flows in biochemical reactions, in particular describing the Faraday-equivalent potential approach to
modelling electrochemical phenomena, and we describe a modified mass action
kinetics approach which will be central to development of a simplified
thermodynamic modelling approach.
A set of Python based tools has been developed to assist the
development and analysis of bond graph models and these tools are briefly outlined.

We then discuss
%\S~\ref{sec:mitoch-electr-transp} discusses 
the Mitochondrial Electron Transport Chain (ETC) as an example of a complex biomolecular system which can be successfully modelled by a simple physically-plausible 
model, and use  
%\S~\ref{sec:model-fitting} uses 
data from  \citet{BazBeaVin16} to derive parameters of the physically-plausible of the ETC to show that this simple model behaves the same as a fully mechanistic description of the ETC. 
%\S~\ref{sec:conclusion} 
Finally we conclude with suggestions for future
research directions using simplified physically-plausible modelling as a strategy in systems and synthetic biology. 

\section{Modelling Bioenergetics of Biochemical Systems using Bond Graphs}
\label{sec:bond-graph-approach}

Bond graphs provide a convenient modular framework for modelling energy flow within and across different physical domains: electrical, mechanical, chemical and so on; and as such are useful for representing biomolecular systems. In brief, bonds represent pairs of variables: potential and flow, whose product is power. 
In the biomolecular domain, the product of chemical potential $\mu$ (with units \si{\joule\per\mole}) and molar flow $v$ (with units \si{\mole\per\second}) is power with units \si{\joule\per\second} \citep{OstPerKat71,OstPerKat73,GawCra14}.
Bonds connect components which represent either storage or dissipation of energy. In biomolecular systems, chemical potential is stored as concentration of chemical species, denoted \Ce{}
\footnote{In this paper, \Ce components are used to represent chemical
  species and \C components to represent electrical capacitors.}
, whereas chemical reactions, denoted \Re{},  in which chemical species are converted from one form to another are dissipative processes. The biochemical network stoichiometry is represented in the coupling of \Ce{} components via the reactions \Re{} using bonds which represent the flow of energy, connected using common potential \zero (`zero') and common flow \one (`one') junctions. 

\begin{figure}[tbp]
  \centering
  \Fig{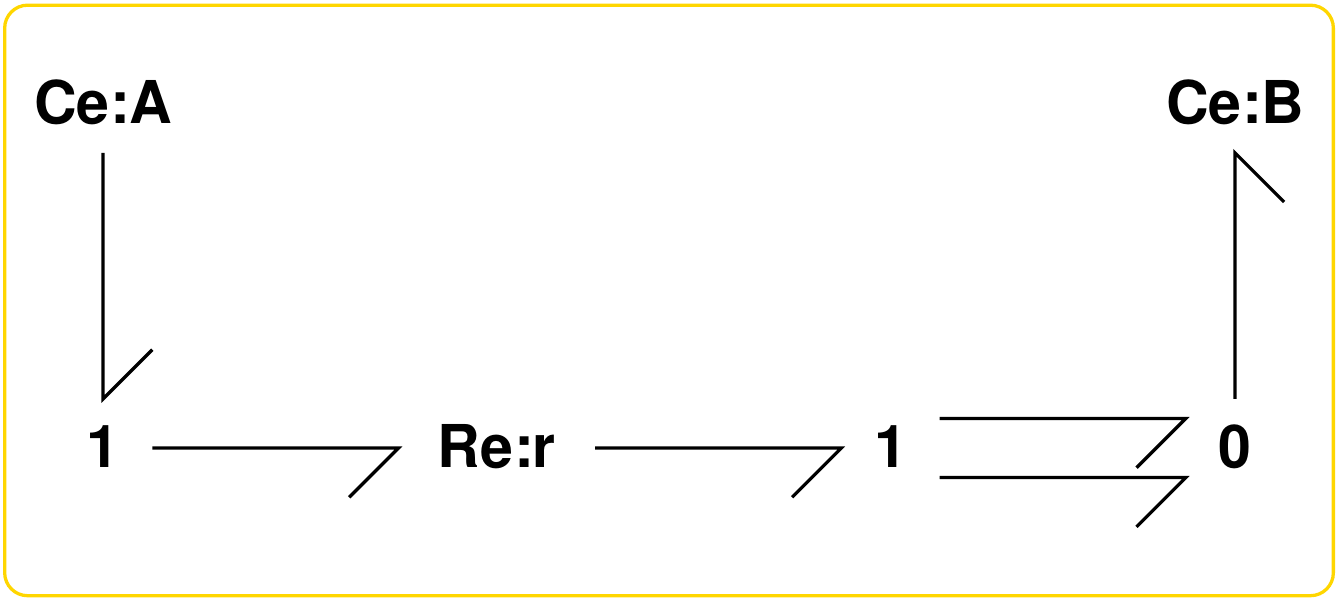}{0.5}
  \caption{Bond Graph representation of \ch{A <>[ r ] 2 B}.
    The bond graph  components \BCe{A} and  \BCe{B} represent species
    \ch{A} and \ch{B}; the bond graph  component \BRe{r} represents
    the reaction the bonds $\rightharpoondown$ together with the zero
    \zero and one \one junctions define the
    stoichiometry~\citep{GawCra14}. The bonds carry the energy
    covariables chemical potential $\mu$ and and
    molar flow $v$.
}
  \label{fig:ABB}
\end{figure}
%As discussed previously \citep{OstPerKat71,GawCra14,Gaw17a}, 
To illustrate, Figure \ref{fig:ABB} is the bond graph representation of the chemical
reaction:
\begin{equation}
  \label{eq:ABB_reac}
  \ch{A <>[ r ] 2 B}
\end{equation}
%In short, \BCe{A} and \BCe{B}% components represent the chemical species \ch{A} and \ch{B} respectively and \BRe{r} the corresponding reaction.
%
\Ce{} components correspond to constitutive relations which relate the chemical potential to the amount of chemical species stored: the constitutive relations of \BCe{A} and \BCe{B} are:
\begin{align}
  \mu_A &= RT \ln K_A x_A\label{eq:CR_A0}\\
  \mu_B &= RT \ln K_B x_B\label{eq:CR_B0}
\end{align}
where $x_A$ and $x_B$ are the concentrations of \ch{A} and \ch{B}, $K_A$ and $K_B$ are species thermodynamic constants (\si{\per\mole}) for \ch{A} and \ch{B}, specific to each chemical species, $R$ is the universal gas constant and $T$ the absolute temperature.

The constitutive relations for the reaction components \Re{} provide the relationship between forward and backward chemical affinities $A$ (stoichiometric combinations of the chemical potentials) which provide the driving force for the reaction, and the molar flow (the reaction rate) $f$. The stoichiometry of reaction \BRe{r} with formation of 2 molecules of species $B$ for each molecule of $A$ is represented by the two parallel bonds on the right hand side of \BRe{r}.
With mass-action kinetics, the constitutive relation of \BRe{r} is
\begin{equation}
  f = \kappa \lb \exp \frac{A^f}{RT} - \exp
      \frac{A^r}{RT} \rb \label{eq:MA}
      \end{equation}
where $\kappa$ is a reaction rate constant (\si{\mole\per\second}), specific to each reaction, and the forwards and backwards affinities are given by
\begin{align}
  A^f &= \mu_A\\
  \text{and }
  A^r &= 2\mu_B
\end{align}
Combining these expressions gives the familiar mass-action expression for the reaction flow $f$:
\begin{equation}
  \label{eq:ABB_v}
  f = \kappa \lb K_Ax_A -  K_B^2x_B^2 \rb
\end{equation}
where the forward reaction rate constant $k^+ = \kappa K_A$ and the reverse reaction rate constant $k^- = \kappa K_B^2$.  

% \subsection{Stoichiometry}
% \label{sec:stoichiometry}
The bond graph approach is naturally allied to stoichiometric concepts \citep{KliLieWie16,Pal06,Pal11,Pal15}. In particular,
the stoichiometric matrix $N$ can be automatically generated from the
network represented in the system bond graph. $N$ can be used to give species flows $f_x$ in terms of reaction flows $f$ and, conversely, reaction affinity $A$ in terms of
species potentials $\mu$:
\begin{xalignat}{2}\label{eq:N}
  f_x &= N f& A &= -N^T \mu
\end{xalignat}
In the case of the system of Figure \ref{fig:ABB}:
\begin{equation}
  N =
  \begin{pmatrix}
    -1&2
  \end{pmatrix}^T
\end{equation}

\subsection{Modified mass action kinetics}
\label{sec:modified-mass-action}
Simplified representation of biomolecular system requires a representation of the reaction network that approximates, but does not fully represent the complete set of biochemical reactions. Physically-plausible models of biomolecular systems will therefore typically contain reactions which are an approximation to a sequence of elementary reactions. Thus even if elementary reaction steps have the mass action
kinetics of Equation \eqref{eq:MA}, this would not necessarily be the
case for the overall reactions used to represent the system \citep[see for example][chapter 17]{AtkPauKee18}.

In bond graph terms, one may represent the dissipative reaction
component with any appropriate constitutive relation for the reaction
flow $f$ in terms of the forward and reverse affinities: mass action,
as given by \eqref{eq:MA} leading to \eqref{eq:ABB_v} is one
example. In particular, non-elementary reactions may be represented
using rate equations where, unlike the mass-action formulation, the
concentration exponents are not the stoichiometric coefficients. One
particular case of this would be to divide all of the stoichiometric
coefficients by an positive integer constant $\alpha$ in the rate
equations.
Thus, for example, if $\alpha=2$, the reaction rate \eqref{eq:ABB_v} corresponding to
the reaction \eqref{eq:ABB_reac} would become:
\begin{equation}\label{eq:ABB_reac_2}
  f = \kappa \lb \sqrt{K_Ax_A} - K_Bx_B\rb
\end{equation}
This corresponds to the non-integer stoichiometry
\begin{equation}
  N_\alpha =
  \begin{pmatrix}
    -\frac{1}{2}&1
  \end{pmatrix}^T
\end{equation}
Note that in the context of modelling the Mitochondrial Electron Transport Chain, the exponent $1/2$, corresponding to $\alpha=2$, commonly appears in the
flux expression for complex III, as given by \citet[equation B72]{Bea05} and
\citet[equation 7.38]{BeaQia10} for example, and 
the exponent $1/4$, corresponding to $\alpha=4$, appears in the
flux expression for complex IV given by \citet[equation B73]{Bea05} and
\citet[equation 7.41]{BeaQia10}. This can be achieved by replacing the mass-action
formula \eqref{eq:MA} by the \emph{modified mass-action} (MMA) formula:
\begin{align}
  f &= \kappa \lb \exp \frac{A^f}{\alpha RT} - \exp
      \frac{A^r}{\alpha RT} \rb\label{eq:MMA0}
\end{align}
which contains the additional parameter $\alpha$, which is used below
%\S~\ref{sec:model-fitting} 
as an essential part of the model fitting process.

For thermodynamic consistency, it is important that Equation
\eqref{eq:MMA0} represents a dissipative system; that is, any non zero
flow dissipates energy \citep{Wil72,PolWil97,Wil07a}. With this in
mind, it is now shown that the MMA equation can be rewritten in
mass-action form but with the positive \emph{constant} $\kappa$ replaced by
the positive \emph{function} of concentration $\kappa_\alpha$.
As a simple example of this, it can be verified that Equation
\eqref{eq:ABB_reac_2} can be rewritten as
\begin{align}
  f &= \kappa_\alpha \left(K_Ax_A - K_B^2x_B^2\right)\label{eq:ABB_reac_2_alt}\\
  \text{where }
  \kappa_\alpha(x_A,x_B) &= \frac{\bar{\kappa}}{\sqrt{K_Ax_A} + K_Bx_B}\label{eq:kappa_alpha}
\end{align}
As $x_A$ and $x_B$ are positive, $\kappa_\alpha$ is also
positive. Thus \eqref{eq:ABB_reac_2_alt} corresponds to the mass-action
equation \eqref{eq:MA} with the positive constant $\kappa$ replaced by
the positive function of concentration $\kappa_\alpha(x_A,x_B)$.
% As shown
% in
% %%Appendix%%
%%\ref{sec:modified-mass-action-1},
The general modified-mass
action kinetics of Equation \eqref{eq:MMA0} can also be rewritten in
mass-action form with the positive constant $\kappa$ replaced by
the positive  $\kappa_\alpha$:
\begin{align}
  f &= \kappa_\alpha(A^f,A^r)  \lb \exp \frac{A^f}{RT} - \exp
      \frac{A^r}{RT} \rb\label{eq:MMA_alpha}
\end{align}

\subsection{Redox reactions}
\label{sec:redox}

%``Whereas all redox reactions can quite properly be described in thermodynamic terms by their Gibbs energy changes, electrochemical parameters can be employed because the reactions involve the transfer of electrons.''~\citep[\S~3.3]{NicFer13}. 
%
Oxidative phosphorylation involves a series of electrochemical redox reactions. \citeauthor{NicFer13} write that `Whereas all redox reactions can quite properly be described in thermodynamic terms by their Gibbs energy changes, electrochemical parameters can be employed because the reactions involve the transfer of electrons.''~\citep[chapter 3.3]{NicFer13}. 
Rather than to deal directly with conversion between electrical and chemical potentials and associated variables, it is convenient to have a common system of units and
convert the chemical energy covariables chemical potential and molar
flow to equivalent electrical energy covariables voltage and current \citep{Gaw17a}.
The relevant conversion factor is \emph{Faraday's constant}
$F\approx\SI{96485}{C.mol^{-1}}$ \citep{NicFer13,GawSieKam17,Gaw17a}.
In particular, we define:
\begin{xalignat}{2}
  &\text{Faraday-equivalent potential}& \phi &= \frac{\mu}{F}~\si{(\volt)}\label{eq:phi}\\
  &\text{Faraday-equivalent flow}& f &= F v~\si{(\ampere)}\label{eq:f}
\end{xalignat}
Using these Faraday-equivalent variables, the \Ce constitutive relations \eqref{eq:CR_A0} and
\eqref{eq:CR_B0} become:
\begin{align}
  \phi_A &= V_N \ln K_A x_A\label{eq:CR_A}\\
  \phi_B &= V_N \ln K_B x_B\label{eq:CR_B}\\
  \text{where }
  V_N &= \frac{RT}{F} \approx
        \SI{26}{\milli\volt}\label{eq:V_N}
\end{align}
and the modified
mass-action formula \eqref{eq:MMA0} becomes:
\begin{align}
  f &= \kappa \lb \exp \frac{A^f}{\alpha V_N} - \exp
      \frac{A^r}{\alpha V_N} \rb\label{eq:MMA}
\end{align}

As noted by \citeauthor{NicFer13}, an advantage of transforming the chemical potentials into equivalent electrical potentials in the treatment of redox reactions is: ``the ability to dissect the overall electron transfer into two half-reactions involving the donation and acceptance of electrons, respectively.''~\citep[chapter 3.3]{NicFer13}.
For example, the two half-reactions:
\begin{xalignat}{2}
  \label{eq:Redox}
  \ch{A &<> [ r1 ] C + 2 e1-} &
  \ch{B + e2- &<> [ r2 ] D}
\end{xalignat}
electron \ch{e1-} donation in the first (oxidation of A), and \ch{e2-}
acceptance in the second (reduction of B), correspond to the overall
reaction:
\begin{equation}
  \label{eq:ABCD}
  \ch{A + 2 B  <> [ r ] C + 2 D }
\end{equation}

\begin{figure}[tbp]
  \centering
  \Fig{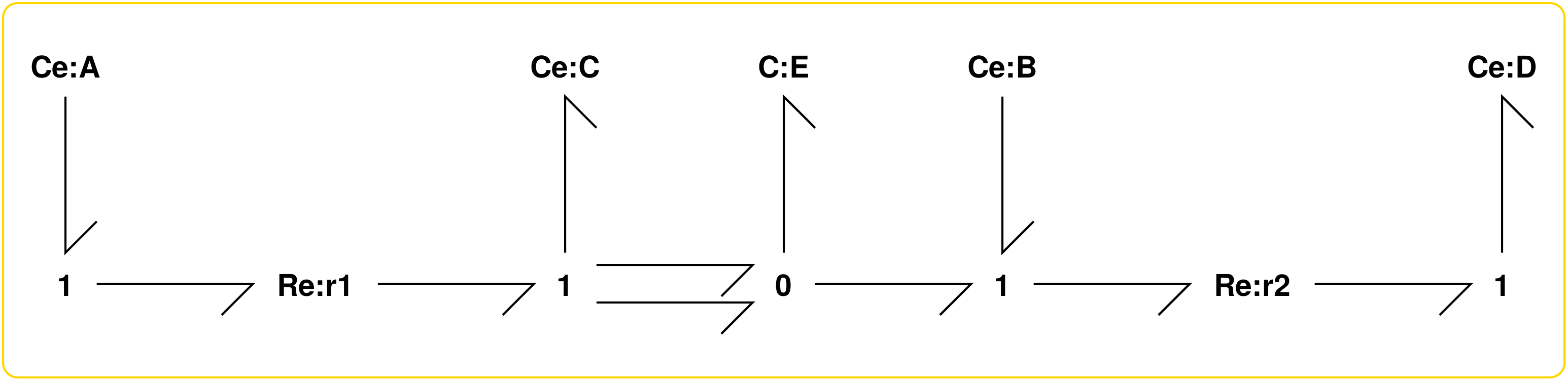}{0.9}
  \caption{Bond Graph representation of two half-reactions:
    \ch{A <> [ r1 ] C + 2 e-} and \ch{B + e- <> [ r2 ] D}.
    As in Figure \ref{fig:ABB}, the bond graph components \BCe{A},
    \BCe{B}, \BCe{C} and
    \BCe{D} represent species \ch{A}, \ch{B}, \ch{C} and \ch{D}; the bond graph
    component \BRe{r1} represent the two half-reactions; the bonds
    $\rightharpoondown$ together with the zero \zero and one \one
    junctions define the stoichiometry.
    The component \BC{E} represents the electrons of the half-reaction
    and is related to the redox potentials. The corresponding voltage
    is $V$.
    As discussed in \S~\ref{sec:redox}, the bonds carry the energy
    covariables electrical potential $\phi$ and and current $v$.  }
  \label{fig:Redox}
\end{figure}

Figure \ref{fig:Redox} shows a bond graph representation of these two half-reactions \eqref{eq:Redox}, explicitly representing the transfer of electrons \ch{e-} using the linear electrical capacitor represented by \BC{E} with voltage $V$; the two-electron stoichiometry of reaction \ch{r1} is represented by the two parallel bonds.
If reaction \ch{r1} is in equilibrium, then the voltage $V$ is exactly
that required to stop reaction \ch{r1} from proceeding and thus
$V=-E_1$ where $E_1$ is the \emph{redox potential} of reaction
\ch{r1}. Conversely, if reaction \ch{r2} is in equilibrium, then the
voltage $V$ is exactly that require to stop reaction \ch{r2} from
proceeding and thus $V=-E_2$ where $E_2$ is the \emph{redox potential}
of reaction \ch{r2}.

% The redox potential $E$ of half-reactions is related to the Gibbs'
% free energy $\Delta G$ of the half-reaction \citep{NicFer13}. In
% particular
% \begin{equation}\label{eq:dG}
%   \Delta G = -nFE
% \end{equation}
% where $n$ is the number of electrons and $F$ is the Faraday constant.
%

\begin{New}
\subsection{Hierarchical Modelling}\label{sec:hier-modell}
Hierarchical modelling and modularity provide one approach to
understanding the complex systems associated with cellular biochemistry
\citep{HarHop99,Lau00,CseDoy02,BruWesHoe02,BruSnoWes08,SzaPerSte10}.
Bond graphs provide an effective foundation for modular construction
of hierarchical models of biochemical systems
\citep{GawCurCra15,GawCra16}.
Bond graphs model the interaction between modules, in particular retroactivity
\citep{JayVec11,Vec13,VecMur14}, in a straightforward manner whilst
retaining thermodynamic compliance.

Bond graph modules use the notion of \emph{chemostats}
\citep{PolEsp14,GawCra16} which have an number of interpretations:
\begin{enumerate}
\item \label{item:1} one or more species is fixed to give a constant concentration; this implies that an appropriate external
  flow is applied to balance the internal flow of the species.
\item \label{item:2} as a \Ce component with a fixed state.
\item \label{item:3} as a module \emph{port} through which chemical, mechanical or electrical energy flows.
\end{enumerate}
Thus if the bond graph of Figure \ref{fig:Redox} were to be used as a
module, then \BCe{A}, \BCe{B}, \BCe{C} and \BCe{D} could be
chemostats. If the module were to be examined in isolation, then the
interpretations of items \ref{item:1} and \ref{item:2} would be used;
if, on the other hand, the module were to be embedded in a larger
system, then the interpretation of item \ref{item:3} would be used.

When examining the properties of a complex system, such as a whole
cell model, the replacement of some modules by physically-plausible
equivalents with the same ports would not only reduce computational
complexity but also allow attention to be focussed on detailed models
of other modules.

\end{New}
\begin{New}
\subsection{Dynamical Simulation}
Bond graphs, together with the component constitutive relationships,
can be used to automatically derive the ordinary differential
equations (ODE) describing the system dynamics \citep{KarMarRos12}. These ODEs
can be in symbolic form or in the form of computer code for a
particular simulation engine.
In general, a set of ODEs does not guarantee thermodynamic
consistency; but, because these ODEs are derived from a bond graph,
they inherit the thermodynamic properties of the bond graph.

In some systems, the system states are not independent; in particular,
biomolecular systems usually have conserved moieties. In such
circumstances, the system bond graph can be used to automatically
generate the minimal number of ODEs describing the independent states
from which dependent states and reaction flows can be derived
\citep[\S~3(c)]{GawCra14}.

Although in this paper we have focused on species described by the
standard logarithmic constitutive relation \eqref{eq:CR_A0} and
reaction flows determined by the mass action formula \eqref{eq:MA},
bond graph components can have a wide range of constitutive relations
constrained only by thermodynamics. A simple example of this is the
modified mass-action formula of equation \eqref{eq:MMA0}; a more
complex example is the Goldman-Hodgkin-Katz flux equation used in bond
graph models of action potential \citep{GawSieKam17}.
Moreover, models with complex characteristics, such as transporters,
can be built from simple bond graph  components \citep{PanGawTra19}.
\end{New}

\subsection{\BGT -- a Python Toolkit}
\label{sec:bond-graph-tools:}
Computational tools are necessary for model capture, parameterisation
and simulation. As the name suggests, \BGT is an application
programming interface (API) for capturing, simplifying and simulating
bond graph models, and is an important part of the bond graph approach
\citep{CudGawPanCra19X}.

\BGT is written in Python and is built upon the Scientific Python
(SciPy) libraries, all of which are open source and easily
accessible. The core use-case of \BGT is to turn bond graphs into a
set of reduced equations which can be then passed into other SciPy
libraries (parameter estimation routines, or ODE integrators, for example).  As model
reduction is performed symbolically, the simplification routines are
free from numerical errors, which is important for systems involving
parameters that are unknown.

%% Appendix%%
% \ref{sec:using-bgt} briefly describes how \BGT can be used to build a
% model of the chemical system shown in Figure \ref{fig:Redox}.

\section{A Simplified Physically-Plausible Model for the Mitochondrial Electron Transport Chain}
\label{sec:mitoch-electr-transp}

%%As discussed in more detail by \citet{Gaw17a},
Mitochondria make use of redox reactions to provide the power driving
many living systems. The key process in the generation of ATP is
\emph{chemiosmotic} energy transduction, whereby a sequence of redox
reactions pumps protons across the mitochondrial inner membrane to
generate the \emph{proton-motive force} (PMF), an electrochemical
gradient which is then used to power the synthesis of ATP. Generation
of the PMF is accomplished by the mitochondrial electron transport
chain. Beard and colleagues have developed the most comprehensive
thermodynamically consistent models of mitochondrial oxidative
phosphorylation including the electron transport chain
\citep{Bea05,WuYanVin07,BazBeaVin16}. Recently \citet{Gaw17a} provided
a bond graph model of mitochondrial oxidative phosphorylation based
on the redox reactions associated with complexes \ch{CI}, \ch{CIII}
and \ch{CIV} of the mitochondrial electron transport chain.

\begin{New}
In contrast, here we develop a simple, but physically-plausible,
model based on the overall chemical reaction of the Electron Transport
Chain in which \ch{2 NADH} is combined with \ch{O2} and \ch{2 H+} to
give \ch{2 NAD+} and \ch{2 H2O}; the 2 protons (\ch{2 H+}) are consumed
from the mitochondrial matrix.
The free energy of this overall reaction pumps 20 protons across the
mitochondrial inner membrane.
Denoting the protons in the mitochondrial matrix as \ch{H_x+} and
those in the mitochondrial inner membrane space as \ch{H_i+} the
overall reaction is thus:
\begin{equation}\label{eq:ETC_reac}
  \ch{2 NADH + O2 + 22 H_x+ <> 2 NAD+ + 2 H2O + 20 H_i+}
\end{equation}  
Reaction \eqref{eq:ETC_reac} can be rewritten as the weighted sum of
three reactions:
\begin{xalignat}{2}
  \ch{NADH &<>[ r1 ] NAD+ + H_x+ + 2 e1-}&(\times 2)\label{eq:donate}\\
  \ch{O2 + 4 H_x+ + 4 e2- &<>[ r2 ] 2 H2O}&(\times 1)\label{eq:consume}\\
  \ch{5 H_x+ + e1- &<>[ r_{loss} ] 5 H_i+ + e2-}&(\times 4)\label{eq:pump}
\end{xalignat}
\end{New}
\begin{New}
  \noindent
  where \ch{e1-} and \ch{e2-} are the electrons transfered from and to
  the left and right half reactions respectively as discussed in \S~\ref{sec:redox}.
\end{New}
Reaction \eqref{eq:donate} converts \ch{NADH} to \ch{NAD} producing a
proton \ch{H_x+} in the mitochondrial matrix and
\emph{donating} two electrons \ch{e1-}. 
%it corresponds to the half-reaction \eqref{eq:NADH} discussed in \S~\ref{sec:nominal-conditions}.
%
\begin{New}
Reaction \eqref{eq:consume} converts \ch{O2} and protons
\ch{H_x+} in the mitochondrial matrix and
\emph{consumes} four electrons \ch{e2-} to produce water \ch{H2O}. 
\end{New}
Reaction \eqref{eq:pump} \emph{transfers} electrons \ch{e1-} to
\ch{e2-} and, in so doing, utilises the corresponding free energy to
pump five protons from the mitochondrial matrix \ch{H_x+} to the
mitochondrial inter-membrane space \ch{H_i+} against the \ch{H+}
concentration gradient and the trans-membrane electrical potential
$\Delta \Psi$.

\begin{figure}[tbp]
  \centering
  \begin{New}
  \Fig{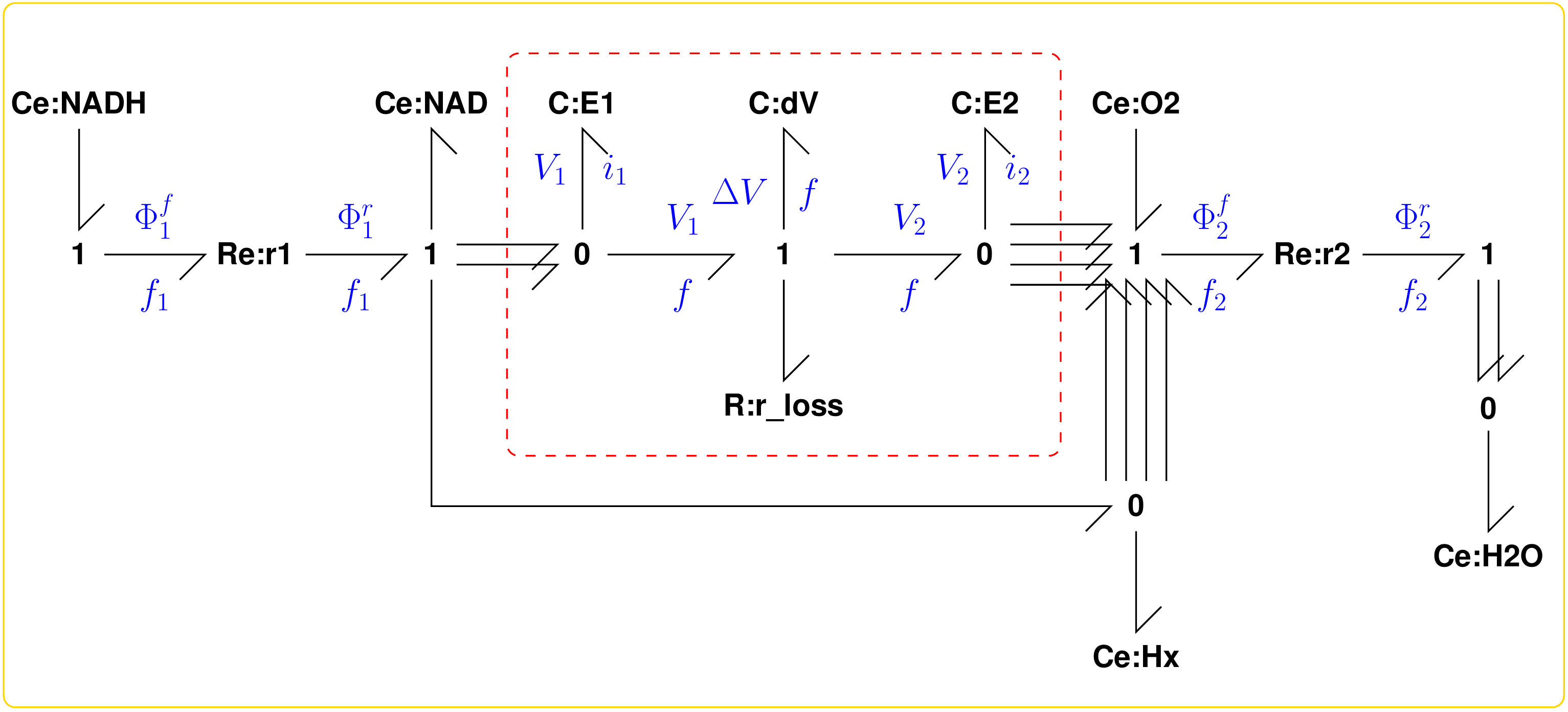}{0.9}
\end{New}
\caption{Mitochondrial electron transport chain: a
    physically-plausible model.
    The reaction represented by \BRe{r1} is the electron donating
    reaction \eqref{eq:donate} and the reaction represented by
    \BRe{r2} is the electron consuming reaction \eqref{eq:consume}.
    The dashed box demarcates the electrical part of the model: the
    electrical resistor \BR{r\_loss} models electrical energy
    dissipation; \BC{E1} and \BC{E2} are electrical capacitors
    accumulating donated and consumed electrons.
    The electrical capacitor \BC{dV} corresponds to the net voltage
    available to pump protons across the mitochondrial inner membrane.
  }
  \label{fig:ETC_abg}
\end{figure}
Figure \ref{fig:ETC_abg} shows the bond graph of a
physically-plausible model of the mitochondrial electron transport
chain. The the components of the model are:
\begin{enumerate}
\item The electron donation reaction \ch{r1} \eqref{eq:donate}  is
  represented by \BRe{r1}  and the associated species by \BCe{NADH},
  \BCe{NAD} and \BCe{Hx}. The electrons \ch{e1-} accumulate in the
  electrical capacitor \BC{E1}.
\begin{New}
\item The electron consumption reaction \ch{r2} \eqref{eq:consume}  is
  represented by \BRe{r2}  and the associated species by \BCe{O2},
  \BCe{H2O} and \BCe{Hx}. The electrons \ch{e2-} accumulate in the
  electrical capacitor \BC{E2}.
\end{New}
\item The electron transfer part of the electron transfer/proton pump
  \eqref{eq:pump} is modelled by the two electrical capacitors \BC{E1}
  and \BC{E1} and the (linear) electrical resistor (with resistance
  $r_{loss}$) \BR{r\_loss}. The voltage $V_1$ associated with \BC{E1}
  is the redox potential of half reaction \eqref{eq:donate}
  and the voltage $V_2$ associated with \BC{E2} is the redox potential
  of half reaction \eqref{eq:consume}.
  The electrical capacitor \BC{dV} with voltage $\Delta V$ represents
  by the the net redox potential minus the potential drop associated
  with the resistor $r_{loss}$:
  \begin{equation}
    \label{eq:elect}
    \Delta V = V_1 - V_2 - r_{loss}f
  \end{equation}
\item Proton transfer is not explicitly modelled in
  Fig.~\ref{fig:ETC_abg}. However, the corresponding membrane
  potential $\Delta \Psi$ is given in terms of $\Delta V$ as by:
  \begin{equation}    \label{eq:PMF}
    \Delta \Psi = \frac{\Delta V}{n_p} - \Phi_H
  \end{equation}
  where $n_p= \frac{20}{4} = 5$ is the number of protons pumped per electron and
  $\Phi_H = \phi_{Hi} - \phi_{Hx}$, the chemical potential difference
  due to proton concentration difference across the membrane.
\end{enumerate}

\subsection{Physical Parameters}
\label{sec:nominal-conditions}
\sisetup{table-format=+1.3e2}

The \Ce constitutive relation \eqref{eq:CR_A} can be rewritten in the
alternative form:
\begin{align}
  \phi_A &= \phi^\Std_A + V_N \ln \frac{x_A}{x^\Std_A}\label{eq:CR_alt}
\end{align}
where $V_N$ is given by \eqref{eq:V_N} and $\phi^\Std$ is the
potential of substance \ch{A} at standard conditions where
$x_A = x^\Std_A$.
Using tables of standard chemical potentials $\mu^\Std$, equation
\eqref{eq:phi} can be used to derive the corresponding potential
$\phi^\Std$.
As discussed by \citet{Gaw17a}, the Faraday-equivalent chemical
potential of substance \ch{A} at nominal conditions $\phi^\std$ can be
computed from Faraday-equivalent chemical potential at standard
conditions $\phi^\Std$ from the formula:
\begin{align}
  \phi^\std_A &= \phi^\Std_A + V_N \ln \rho_A\label{eq:nominal}\\
\text{where } \rho_A &= \frac{x_A^\std}{x_A^\Std}\label{eq:rho}
\end{align}
where $x_A^\std$ and $x_A^\Std$ are the concentrations of substance
\ch{A} at nominal conditions and standard conditions.
Table \ref{tab:Physical} shows nominal values $\phi^\std$
for a number of different substances. These values will be used below to model the mitochondrial electron transport chain.
\begin{table}[tbp]
  \centering
  \begin{tabular}{|c|SSS|}
        \hline
    Substance & $\rho$ & $\phi^\Std~\si{(\volt)}$&$\phi^\std~\si{(\volt)}$\\
    \hline
    	\ch{H2O} & 1.000e+00 & -2.443e+00 & -2.443e+00\\
	%%\ch{H^+_i} & 1.318e-07 & 0.000e+00 & -3.729e-01\\
	\ch{H^+_x} & 1.660e-08 & 0.000e+00 & -4.217e-01\\
	\ch{NAD+} & 1.500e-03 & 1.876e-01 & 3.454e-02\\
	\ch{NADH} & 1.500e-03 & 4.074e-01 & 2.544e-01\\
    \ch{O2} & 2.500e-05 & 1.700e-01 & -7.945e-02\\
    \hline
  \end{tabular}
\caption{{Physical Parameters of the Physically-plausible model.}
    $x^\std$ is the concentration at nominal conditions relative to
    standard conditions, $\phi^\Std$ and $\phi^\std$ are the
    Faraday-equivalent potentials at standard and nominal conditions
    related by Equation \eqref{eq:nominal} and where $\rho$ is given by
    Equation \eqref{eq:rho}.
    \ch{H^+_x}are protons in the matrix.
  }\label{tab:Physical}
\end{table}

%
% For example, the half reaction \ch{NADH <> NAD+ + H_x+ + 2 e1-} has
% a net reaction potential at nominal conditions of:
% \begin{equation}
%   \Phi^\std = \phi^\std_{NADH} - \phi^\std_{NAD} - \phi^\std_{Hx}
%   =
%   254.4 - 34.54 - (-421.7) = \SI{641.56}{\milli\volt}
% \end{equation}
% As there are 2 electrons, the redox potential at these conditions is
% \SI{321}{\milli\volt}.
% %
% Similarly, the half reaction \ch{O2 + 4 H_i+ + 4 e2- <> 2 H2O}  has
% a net reaction potential at nominal conditions of:
% \begin{equation}
%   \Phi^\std = \phi^\std_{O2} + 4 \phi^\std_{Hi} - 2 \phi^\std_{H2O}
%   =
%   -79.45 + 4 \times -421.7  - 2 \times -2443
%   = \SI{3120}{\milli\volt}
% \end{equation}
% As there are 4 electrons, the redox potential at these conditions is
% \SI{780}{\milli\volt}.

The stoichiometric equations \eqref{eq:N} can be rewritten in
Faraday-equivalent form as
\begin{xalignat}{2}\label{eq:Nphi}
  f_x &= N f& \Phi &= -N^T \phi
\end{xalignat}
In the case of the half-reaction~\eqref{eq:donate}
\begin{equation}
  \label{eq:NADH}
  \ch{NADH <> NAD+ + H_x+ + 2 e1-}
\end{equation}
the stoichiometric matrix is:
\begin{equation}
  N^T =
  \begin{pmatrix}
    -1&1&1&2
  \end{pmatrix}
\end{equation}
It follows that the reaction potential $\Phi$ is:
\begin{equation}
  \Phi = N^T\phi = \phi^\std_{NADH} - \phi^\std_{NAD} - \phi^\std_{Hx}
  - 2 \phi^\std_{E}
\end{equation}
At equilibrium, $\Phi = 0$ and so:
\begin{align}
  V = \phi^\std_{E} &= \frac{1}{2} \lb \phi^\std_{NADH} - \phi^\std_{NAD} -
                  \phi^\std_{Hx}\rb \notag\\
                &= \frac{1}{2} \lb 254.4 - 34.54 - (-421.7) \rb\approx \SI{320}{\milli\volt}
\end{align}
and so this corresponds to a redox
potential of $E=-V=\SI{-320}{\milli\volt}$ for this half-reaction.

\begin{New}
Similarly, in the case of the half-reaction~\eqref{eq:consume}
\begin{equation}
  \label{eq:O2}
  \ch{O2 + 4 H_x+ + 4 e2- <> 2 H2O}
\end{equation}
% the stoichiometric matrix is:
% \begin{equation}
%   N^T =
%   \begin{pmatrix}
%     -1&-4&-4&2
%   \end{pmatrix}
% \end{equation}
% It follows that the reaction potential $\Phi$ is:
% \begin{equation}
%   \Phi = -\phi^\std_{O2} - 4 \phi^\std_{Hx} - 4 \phi^\std_{E} + 2 \phi^\std_{H2O}
% \end{equation}
% At equilibrium, $\Phi = 0$ and so:
% \begin{align}
%   \phi^\std_{E} &= \frac{1}{4} \lb 2\phi^\std_{H2O} - \phi^\std_{O2} -  4\phi^\std_{Hx}
%                   \rb \notag\\
%                 &= \frac{1}{4} \lb 2 \times -2443 - (-79.5) - 4
%                   \times -421.7 \rb\approx -\SI{780}{\milli\volt}
% \end{align}
the redox
potential is $E=-V=\SI{780}{\milli\volt}$.
\end{New}

%\tbd{Discuss analysis via bond graph tools \url{https://pypi.org/project/BondGraphTools/}.}

\subsection{An explicit formula}
\label{sec:an-explicit-formula}
From a systems point of view, the model of the ETC can be characterised by
the voltage/current relationship 
%at the port represented by the bond graph 
of the bond graph component \BC{dV}. This represents the steady state relationship between the flow (rate of electron transport along the ETC, or equivalently the rate of oxygen consumption) and the mitochondrial membrane potential which is established. Letting $n_1$ and $n_2$ be the number of bonds
connecting reactions \ch{r1} and \ch{r2} to the electrical subsystem,
the steady-state flows are related by:
\begin{equation}
  \label{eq:ETC-flows}
  f = n_1 f_1 = n_2 f_2
\end{equation}
and the steady-state potentials by Equation \eqref{eq:elect} with:
\begin{align}
  V_1  &= \frac{1}{n_1} \Phi_1\\
  \text{and}
  V_2  &= \frac{1}{n_2} \Phi_2
\end{align}
Using the modified mass action formula \eqref{eq:MMA}, the reaction
flows are given by
\begin{align}
  f_1 &= \kappa_1 \lb \exp \frac{\Phif_1}{\alpha V_N}
        - \exp\frac{\Phir_1+n_1V_1}{\alpha V_N} \rb\\
  f_2 &= \kappa_2 \lb \exp \frac{\Phif_2+n_2V_2}{\alpha V_N}
        - \exp \frac{\Phir_2}{\alpha V_N} \rb 
\end{align}
where the parameter $\alpha$ remains to be determined (by fitting to data). 
At equilibrium, the flows are zero and thus:
\begin{align}
  V_1 = V_1^{eq} &= \frac{1}{n_1}\lb \Phif_1 - \Phir_1\rb = \frac{1}{n_1}\Phi_1\\
  V_2 = V_2^{eq} &= \frac{1}{n_2}\lb \Phir_2 - \Phif_2\rb = -\frac{1}{n_2}\Phi_2
\end{align}
Writing $\Delta V_1 = V_1-V_1^{eq}$ and $\Delta V_2 = V_2-V_2^{eq}$ it follows that the flows can be rewritten as:
\begin{xalignat}{2}
  f_1 &= (1-\lambda_1)K_1 &
  f_2 &= (\lambda_2-1)K_2 \label{eq:f_12}
\end{xalignat}
where
\begin{xalignat}{2}
  \lambda_1 &= \exp \frac{n_1\Delta V_1}{\alpha V_N} &
  \lambda_2 &= \exp \frac{n_2\Delta V_2}{\alpha V_N}\label{eq:lambda}\\
  K_1 &= \kappa \exp \frac{\Phif_1}{\alpha V_N} &
  K_2 &= \kappa \exp \frac{\Phir_2}{\alpha V_N}
\end{xalignat}
Hence using \eqref{eq:f_12}
\begin{xalignat}{2}
  \lambda_1 &= 1 - \frac{f}{n_1K_1}&
  \lambda_2 &= 1 + \frac{f}{n_2K_2}
\end{xalignat}
Using \eqref{eq:lambda} and \eqref{eq:elect}
\begin{equation} \label{eq:explicit}
%  \boxed{
    \Delta V = \Delta V^{eq}
    + \frac{\alpha V_N}{n_1}\ln \lb 1 - \frac{f}{n_1K_1}\rb
    - \frac{\alpha V_N}{n_2}\ln \lb 1 + \frac{f}{n_2K_2}\rb
    - r_{loss} f
%  }
\end{equation}
where $\Delta V^{eq} = V_1^{eq}-V_2^{eq}$.
Using the results of \S~\ref{sec:nominal-conditions}
\begin{New}
\begin{equation}
  \Delta V^{eq} = 320 + 780 = \SI{1100}{\milli\volt}
\end{equation}
\end{New}
This formula, derived from the simplified model using modified mass
action, thus provides a voltage/current steady state relationship that
describes the operation of the ETC. %
\begin{New}
Using equation \eqref{eq:PMF} and a value of
$\Phi_H=\SI{25}{\milli\volt}$ this corresponds to an equilibrium
mitochondrial membrane potential $\Delta \Psi_{eq}$ of:
$\Delta \Psi_{eq} = \frac{1100}{5} - 25 = \SI{195}{\milli\volt}$.
\end{New}
\subsection{Model Fitting}
\label{sec:model-fitting}
The simplified, physically plausible model of the electron transport chain derived above contains physical parameters which are known \emph{a-priori},
as well as parameters which are
model-dependent and must be obtained by fitting to relevant data.
In particular, the explicit formula \eqref{eq:explicit} relating flow $f$
to potential difference $\Delta V$ has four known physical parameters:
$\Delta V^{eq}$, $V_N$, $n_1$ and $n_2$ and four unknown parameters: $\alpha$,
$K_1$, $K_2$ and $r_{loss}$.
\citet{BazBeaVin16} develop a detailed physically based mathematical
model of mitochondrial oxidative phosphorylation and ROS generation
and compare the results with \emph{in-vitro} experimental data. These
results are used below to fit the parameters of the
physically-plausible model developed above. 
%of \S~\ref{sec:mitoch-electr-transp}.

%There are two categories of data:
%\begin{enumerate}
%\item data generated from a simulation of a (possibly complex) model
%  of the system that has already been obtained;
%\item data obtained from an experiment.
%\end{enumerate}

%Both of these categories will be illustrated here by using the simulation
%results and experimental data of \citet{BazBeaVin16}.

%% \tbd{table of physical parameters and discussion}
%%\sisetup{table-format=+1.4e+2, output-exponent-marker = \text{e} }

Figure 2B in \citeauthor{BazBeaVin16} shows four datasets: two sets of
simulation results corresponding to two concentrations of inorganic
phosphate \ch{Pi}: $[\ch{Pi}]= \SI{1}{\milli M}$ and
$[\ch{Pi}]= \SI{5}{\milli M}$, and two sets of experimental data
corresponding to these two conditions.
\begin{New}
The maximum value for $V_{O_2}$ (Figure 2B in
\citeauthor{BazBeaVin16}) is \SI{150}{~\nano
  \mole \per min \per U~CS}. This value is used to normalise the
reported flows for the purposes of parameter fitting.
Thus in Figure \ref{fig:fit},  the mitochondrial
membrane potential $\Delta \Psi$~\si{\milli\volt} is plotted against
the normalised rate of oxygen consumption
$f$.
\end{New}
\begin{New}
%% Why we use constant NADH, pH etc.
In ideal circumstances, the experimental data would correspond to an
isolated module where the species concentrations corresponding to each
chemostat in the model (see \S~\ref{sec:hier-modell}) were constant.
This is not the case here as, for example, \ch{NADH} is generated from
the mitochondial TCA (citric acid) cycle and so concentration of
\ch{NADH}  depends on
the flow though the TCA cycle.

For the purposes of illustrating parameter fitting in the context of
physically-plausible models, this lack of isolation is not included;
but would be an interesting topic of future research. 
\end{New}
%% \tbd{Discussion of why \ch{Pi} is not in our model and the
%% consequences.}

\begin{New}
  A further approximation is that inorganic phosphate \ch{Pi} does
  not appear in the physically plausible model developed above
%f \S~\ref{sec:mitoch-electr-transp} 
and therefore it
is not possible to take account of the variation of \ch{Pi} in this
approximate model.
\end{New}
However, the experimental data shown in Figure 2B of
\citeauthor{BazBeaVin16} does not show a strong dependence of this
part of the system on \ch{Pi} and so the lack of dependence of our
simplified model
%of \S~\ref{sec:mitoch-electr-transp} 
on \ch{Pi} is reasonable.
%for such a simple model. 
For this reason, experimental data for both values of
\ch{Pi} are considered for the purposes of model fitting to the raw data.
We note that other components of mitochondrial metabolism not modelled
here, such as ATPase, do depend strongly on \ch{Pi}.

%Parameter fits are shown in Figure \ref{fig:fitK}. 
Known physical parameters $\Phi^\Std$ of Table~\ref{tab:Physical}
are drawn from the supplementary material of \citet{WuYanVin07}, the
concentrations of \ch{NAD} and \ch{NADH} from \citet{BazBeaVin16}, the
pH values from \citet{PorGheZan05} and the \ch{O2} concentration from
\citet{Mur09}.
Parameter fitting was implemented using the Python function within the
\texttt{SciPy.optimize} module \texttt{opt.minimize} with \texttt{method =
  "L-BFGS-B"} and parameter bounds of $10^{-12}$ and $\infty$ as the
parameters are all positive. The cost function $\epsilon$ is the
root-mean-square (RMS) difference between the value of $\Delta \Psi$
predicted by equations \eqref{eq:PMF} and \eqref{eq:explicit} and the
data for each value of (normalised) flow.

\begin{figure}[tbp]
  \centering
\begin{New}
  \SubFig{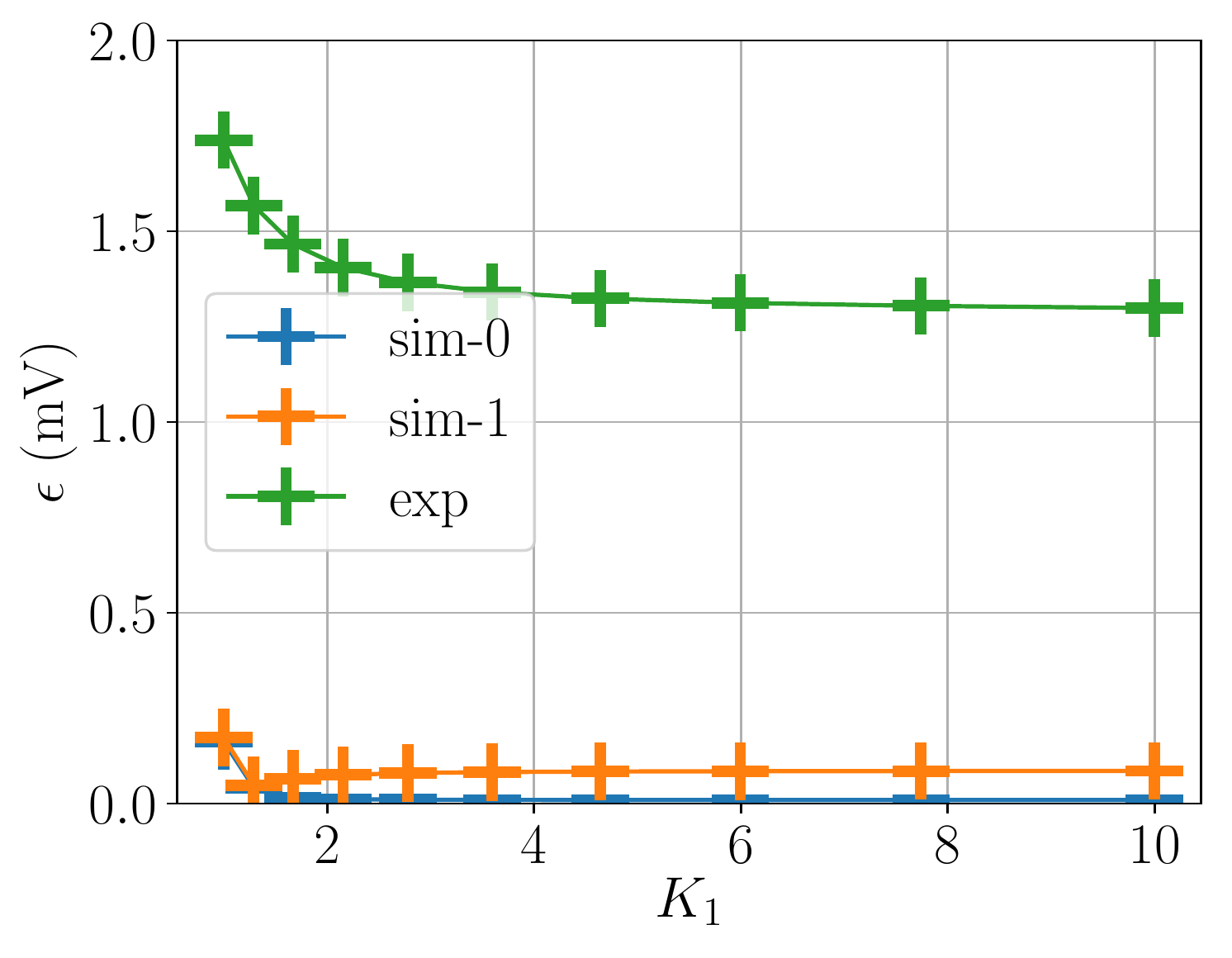}{$K_1$}{0.45}
  \SubFig{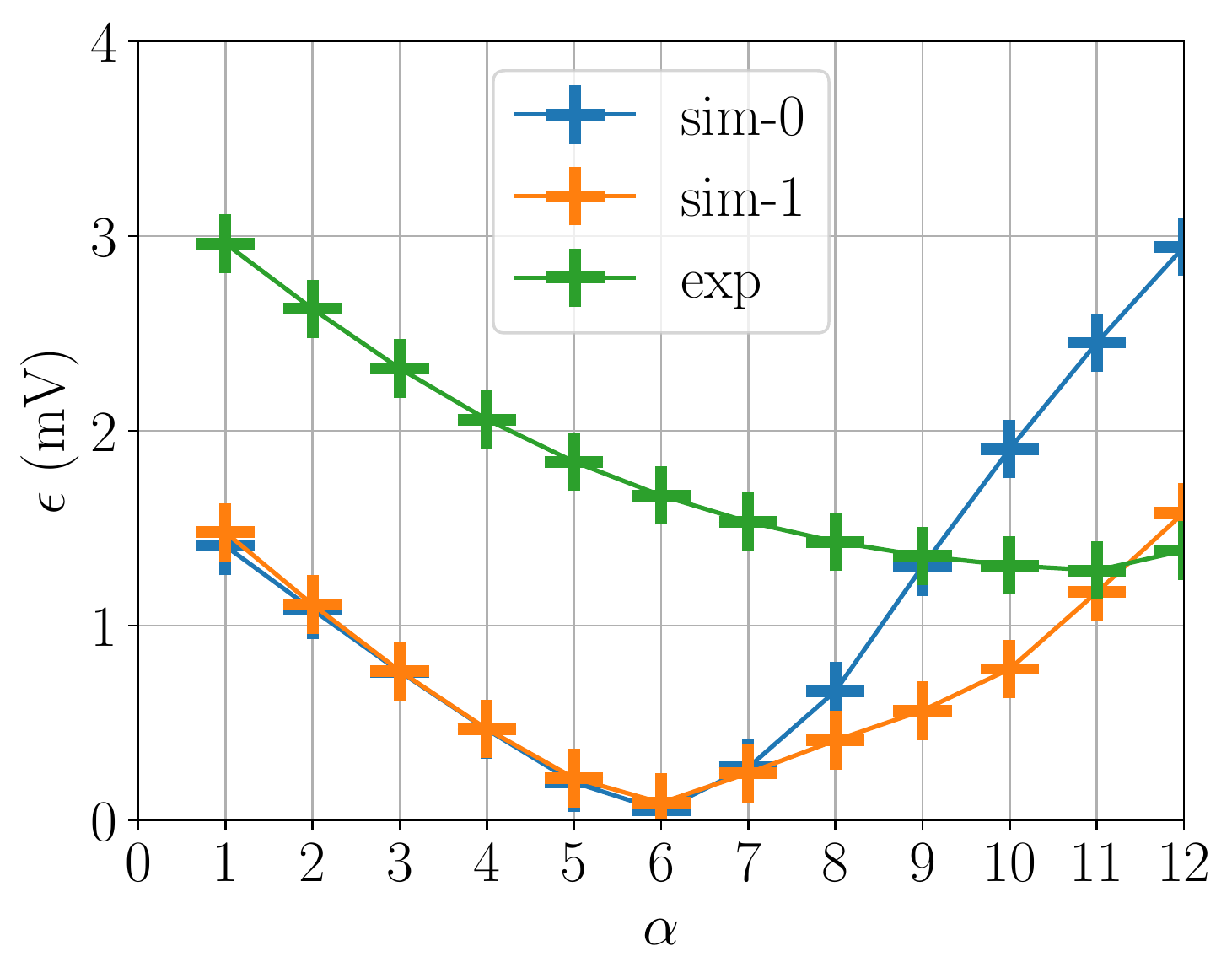}{$\alpha$}{0.45}
\end{New}
\caption{Fitting $K_1$ and $\alpha$.
    (a) The RMS error $\epsilon$ as parameter $K_1$ is varies; all
    other parameters are optimised. The value of $K_1$ is not
    important if greater than about 2; the reaction r1 is in
    equilibrium.
    (b) The RMS error $\epsilon$ as parameter $\alpha$ of the modified
    mass-action kinetics \eqref{eq:MMA} is varies; all other
    parameters are optimised. For data sets \texttt{sim-0} and
    \texttt{sim-1} (data from the experimentally fitted model of
    \citet{BazBeaVin16}) there is a clear minimum at about $\alpha=6$.
  }
  \label{fig:fitK}
\end{figure}

The form of the cost function was examined by fixing one of the four
parameters and minimising with the other three and plotting the cost
against the fixed parameter.
Figure \ref{subfig:errK} shows the dependence of the cost function on parameters $K_1$ and $\alpha$, and indicates that the
parameter $K_1$ has little effect as long as it is greater than
about $K_1=5$. For the rest of this paper, $K_1$ was fixed at a large
positive value and not included further in the optimisation. Thus there are
three significant parameters: $\alpha$, $K_2$ and $r_{loss}$.
Figure \ref{subfig:err} shows that the
parameter $\alpha$ has a significant effect. For the two simulation
data sets, there is a clear minimum at about $\alpha =6$.
The rows 1--3 of Table \ref{tab:results} show the optimal
parameters $\alpha$, $K_2$ and $r_{loss}$, together with the minimal
cost $\epsilon$ for the two simulations and
the experimental data. Rows 4--6 correspond to fixing $\alpha=6$ and
estimating the remaining two parameters $K_2$ and $r_{loss}$.
\sisetup{table-format=+1.1e2}
\begin{table}[tbp]
  \centering
  \begin{New}
  \begin{tabular}{|c|cSSS|}
    \hline
    {Source} & {$\alpha$} & {$K_2$} & {$r_{loss}$ (\si{\milli\ohm})} & {$\epsilon$ (\si{\milli\volt})}\\
    \hline
    $\text{sim}_0$ & 5.8 & 1.8e-03 & 3.1e+01 & 9.2e-03\\
$\text{sim}_1$ & 5.9 & 4.6e-03 & 4.6e+01 & 8.6e-02\\
exp & 10.9 & 1.3e-02 & 1.0e-06 & 1.3e+00\\
\hline
$\text{sim}_0$ & 6.0 & 2.0e-03 & 2.7e+01 & 5.1e-02\\
$\text{sim}_1$ & 6.0 & 4.8e-03 & 4.4e+01 & 9.0e-02\\
exp & 6.0 & 4.4e-03 & 5.6e+01 & 1.7e+00\\
\hline

  \end{tabular}
  \end{New}

  \caption{{Optimal parameters.} The table summarises the fitting
    results using the two sets of simulation data
    ($\text{sim}_0$\&$\text{sim}_1$) and the experimental data (exp) from
    \citet{BazBeaVin16}. Rows 1--3 correspond to free
    $\alpha$ and the rows 4--6 to fixed $\alpha=6$. }
  \label{tab:results}
\end{table}

Finally, the current-voltage relationship derived above and given in \eqref{eq:explicit} is plotted in Figure~\ref{fig:fit} for different sets of fitted parameters given in Table~\ref{tab:results}, along with simulation results from the full model of \citet{BazBeaVin16}, and the corresponding experimental data, showing that the explicit formula for the steady state current/voltage behaviour of the ETC is well captured by the physically plausible model. Figures~\ref{subfig:fit_0_fix}--\ref{subfig:fit_exp_fix} correspond to rows 4--6 of Table \ref{tab:results}; Figure~\ref{subfig:fit_exp_free}
corresponds to row 3 of Table \ref{tab:results}.
\begin{figure}[htbp]
  \centering
  \begin{New}
  \SubFig{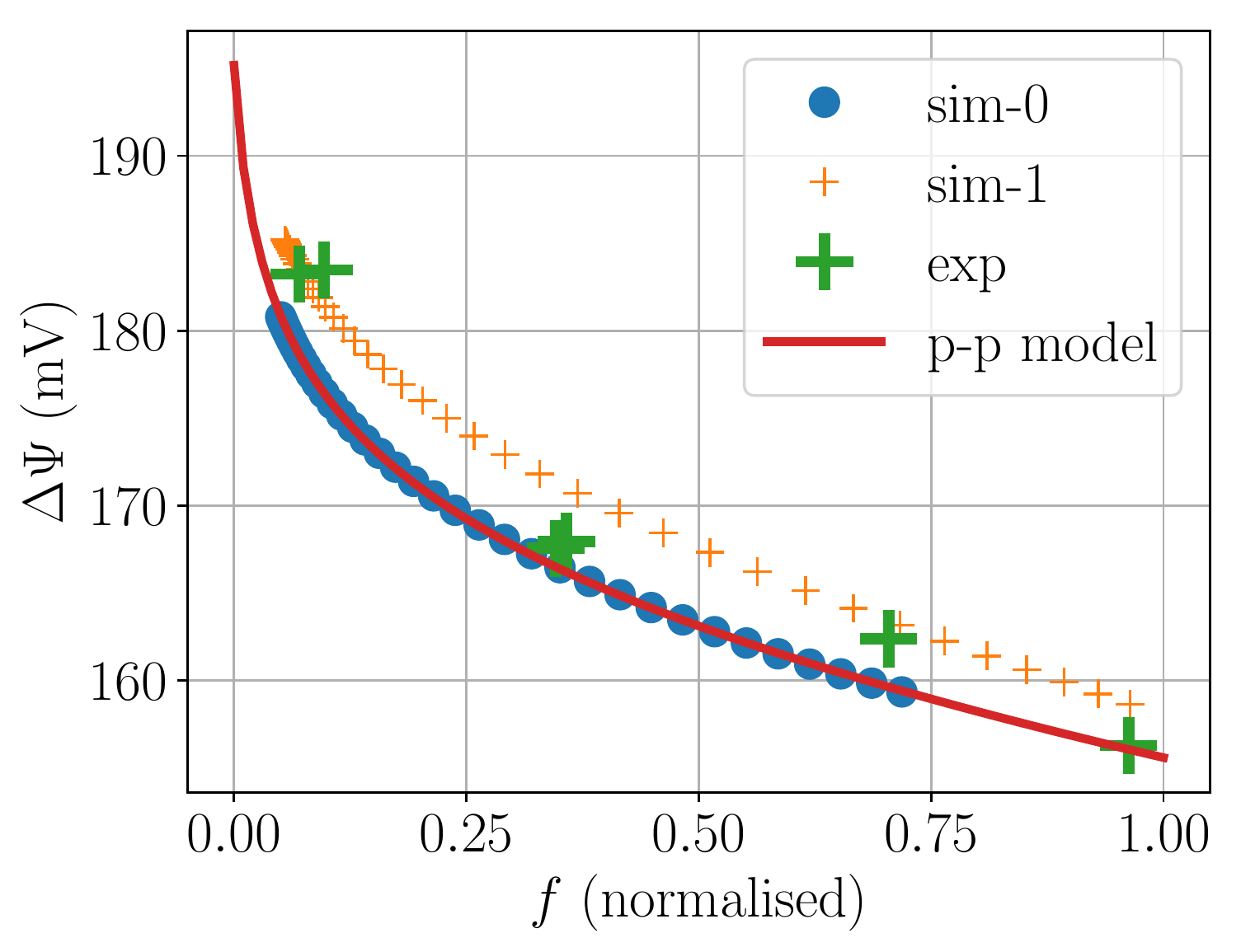}{Fit to $\text{sim}_0$ (fixed $\alpha=6$)}{0.45}
  \SubFig{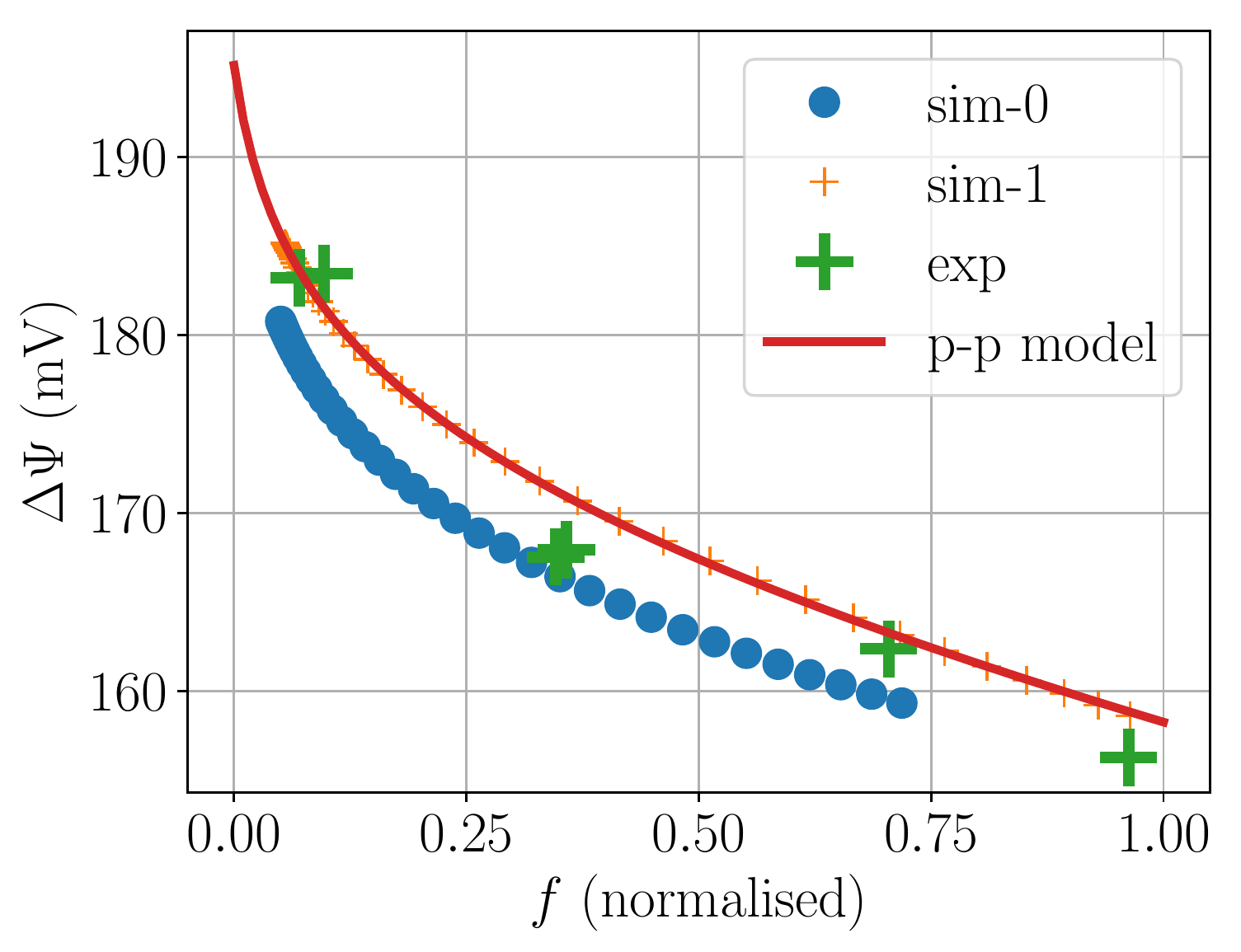}{Fit to $\text{sim}_1$ (fixed $\alpha=6$)}{0.45}
  \SubFig{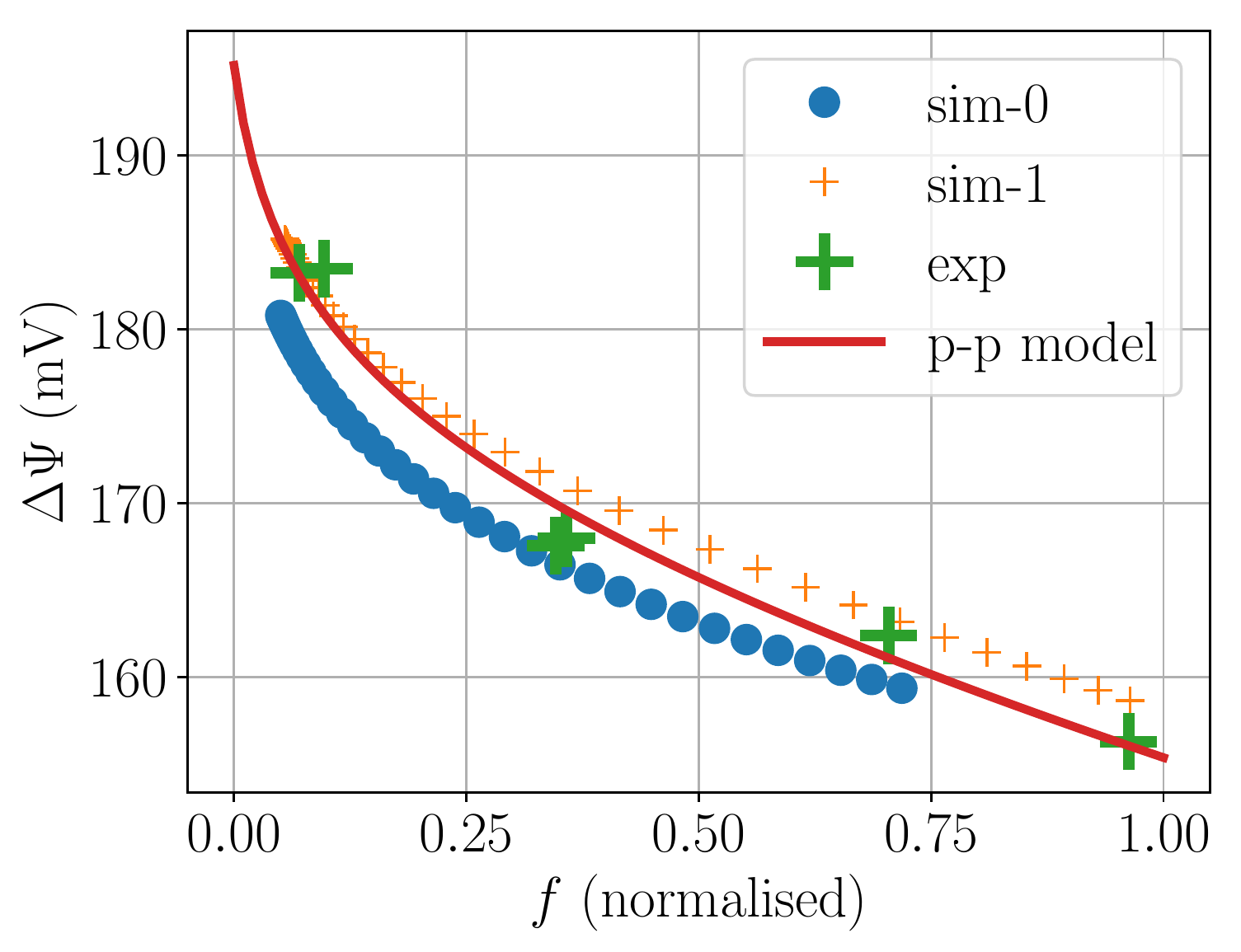}{Fit to exp  (fixed $\alpha=6$)}{0.45}
  \SubFig{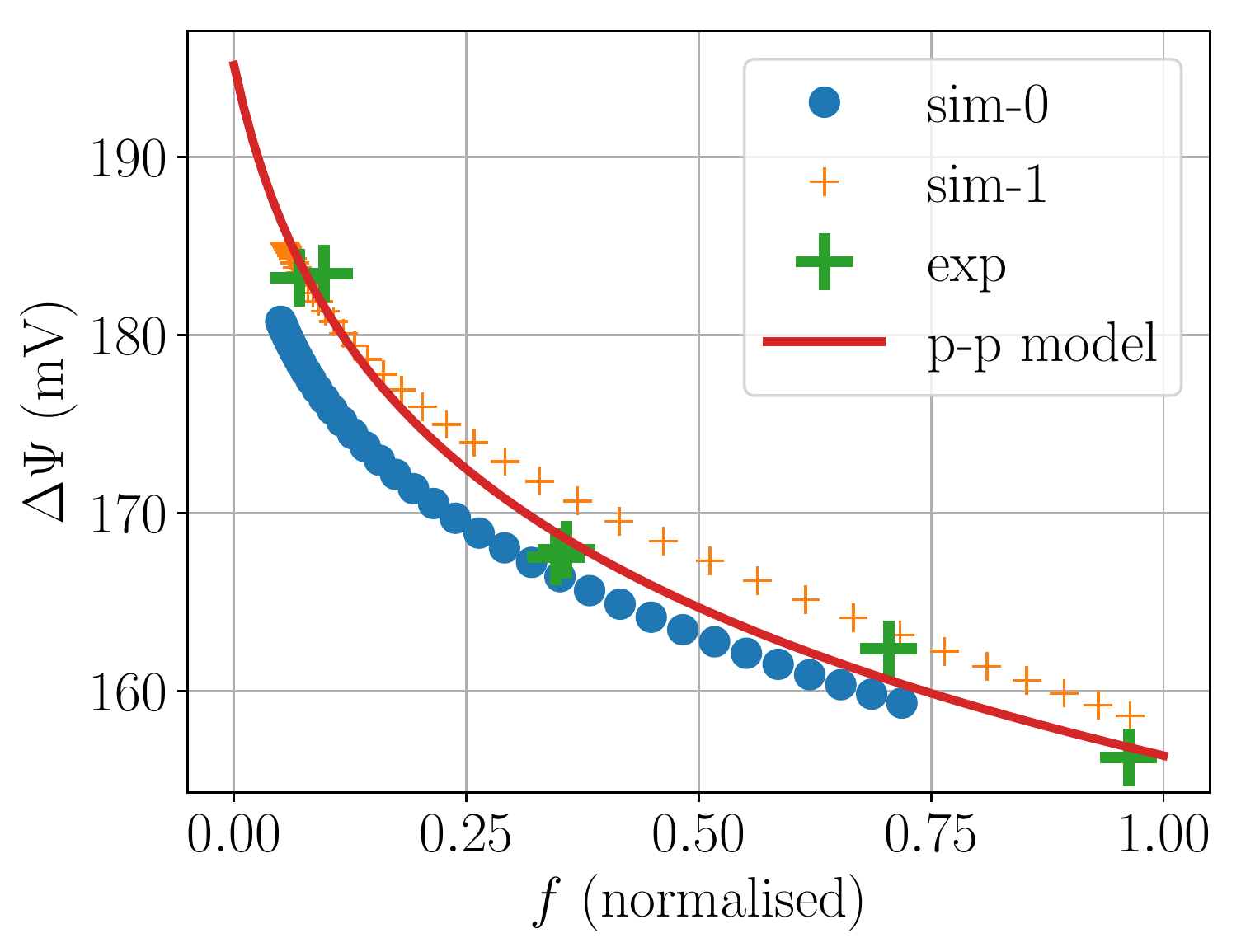}{Fit to exp  (free $\alpha$)}{0.45}
\end{New}
\caption{Model fitting.
    All four plots show four data sets: \texttt{sim-0} and \texttt{sim-1} are simulation
    data from the experimentally fitted simulation of
    \citet{BazBeaVin16} for two values of [\ch{Pi}]; \texttt{exp} is the corresponding
    experimental data; \texttt{pp-model} is the simulation of the
    physically-plausible model using the formula of \S~\ref{sec:an-explicit-formula}.
    (a) The physically-plausible model is fitted to the experimentally
    fitted simulation of \citet{BazBeaVin16} with $[\ch{Pi}]=
    \SI{1}{\milli M}$ using fixed $\alpha=6$
    and estimating $K_2$ and $r_{loss}$. See row 4 of Table
    \ref{tab:results}.
    (b) As (a) but with $[\ch{Pi}]= \SI{5}{\milli M}$. See row 5 of
    Table \ref{tab:results}.
    (c) The physically-plausible model is fitted to the experimental
    points combining both values of $\ch{Pi}$ using fixed $\alpha=6$
    and estimating $K_2$ and $r_{loss}$, see row 6 of Table
    \ref{tab:results}.
    (d) As (c) except that $\alpha$ is also estimated, see row 3 of Table
    \ref{tab:results}.
  }
  \label{fig:fit}
\end{figure}

\newpage
\begin{New}
\section{Discussion}
\label{sec:conclusion}
Simplified models of biochemical and biophysical processes have a central role to play in the development of large whole-cell and multi-scale Physiome models, and the use of such models in biomedical and synthetic biology applications. Here we have argued that such simplified models need to be physically plausible, in the sense that they are consistent with the laws of physics (for example, that they obey mass conservation, are consistent with thermodynamic principles, and so on) as well as providing a suitable fit to available data sets. We have demonstrated that energy-based modelling using bond graphs provides a useful framework for the development of such models. The advantages of thermodynamically-consistent modelling have  been argued recently by us and a number of other authors \citep{BeaQia10,GawCra14,KliLieWie16,GawCra17,MasCov19}. Key amongst these advantages are that models comprised of physically-plausible  components are themselves physically plausible; that such models can be constructed and assembled in a modular fashion; and that such models enforce thermodynamic principles which allow identification of conserved moieties and furthermore restrict the possible parameter space. Here we have shown that by using a modified form of mass action we can generate a simple physically-plausible model of the mitochondrial electron transport chain that is able to reproduce experimentally measured properties of the system. 

%bond graph model is a 'mechanistic' model. 
Bond graphs provide a framework within which is represented both the biochemical network stoichiometry and the constitutive relationship between thermodynamic driving force and biochemical reaction rate for each constituent bond graph element, describing the mechanism of enzymatic processes and so forth. Therefore, bond graphs describe the complete dynamical behaviour of the biochemical system, and can be used to derive the ordinary differential equations describing full system dynamics. 
%exact formula
A particular advantage, however, of a simple model as derived here over a fully mechanistic model is the possibility of deriving explicit formulae for key properties and behaviours of the system. Here we have shown that using the physically plausible modelling approach we can derive an explicit algebraic formula for the flux through the electron transport chain to the PMF that is generated across the mitochondrial membrane at steady state, given in equation~\eqref{eq:explicit}. This is not in general possible from a full dynamical representation (whether represented as a bond graph or otherwise). This is of significance both as it drastically simplifies the model, and also because it allows a direct analysis of the dependence of the mitochondrial membrane potential on parameters of the ETC flux, as discussed below. Such a simplified but thermodynamically realistic representation of mitochondrial energy production is particularly suitable to be used in simulation of spatially-distributed networks of mitochondria \citep{JarGhoDel17,GhoTraCra18}, for which detailed mechanistic 
It remains to be determined how different possible modifications to mass action, or indeed other constitutive relations that may be used to relate chemical potential and reaction rate, affect the ability of simple physically plausible models to represent complex biochemical processes. models of mitochondrial bioenergetics have too high a computational overhead. 

%parameters
The physically plausible model of the electron transport chain derived above contains physical parameters which are known \emph{a-priori}, as well as parameters which are model-dependent and therefore obtained by fitting to relevant data. In particular, the explicit formula \eqref{eq:explicit} relating flow $f$ to potential difference $\Delta V$ has four known physical parameters: $\Delta V^{eq}$, $V_N$, $n_1$ and $n_2$ and four unknown parameters: $\alpha$, $K_1$, $K_2$ and $r_{loss}$.
Using an optimization approach to model fitting, we have shown that the value of the parameter $K_1$ appears unimportant, as long as $K_1>10$; this corresponds to the rate constant $\kappa_1$ of reaction $r_1$ being large enough so that there is negligible potential drop across the reaction; in other words, this requires that the reaction \eqref{eq:donate} is operating essentially at equilibrium under the experimental conditions considered. 

In contrast, the $\alpha$ parameter of the modified mass-action equation \eqref{eq:MMA} is found to be important. Choosing $\alpha=6$ gives a good fit for both the experimentally-fitted simulations and experimental data. This dependence is to be expected as the physically plausible model subsumes a number of individual reactions and this is known to lead to non-stoichiometric exponents~\citep[chapter 17]{AtkPauKee18}.

We have shown that despite it's simplicity, the physically-plausible model fits the (complex) simulation data from \citet{BazBeaVin16} closely (RMS error $\epsilon \ll \SI{1}{\milli\volt}$) for both values of \ch{Pi} and for free $\alpha$ and fixed $\alpha=6$.
Furthermore the experimental data can be fitted with the model with an error $\epsilon$ of about \SI{2}{\milli\volt}.

%Simple vs complex - trade-offs
Despite the success of the simplified, physically-plausible model of the electron transport chain that we have developed here, it should be noted that there are choices and trade-offs inherent in the simplification process. Therefore it is important that such simplified models are analysed and used only in the appropriate context. 
%choice of modification to mass action
Firstly, the simplified model was generated using a specific form of modified mass action kinetics. While motivated by existing literature and approaches for representing non-elemental biochemical reaction steps, different choices could have been made. It remains to be determined how different possible modifications to mass action, or indeed other constitutive relations that may be used to relate chemical potential and reaction rate, affect the ability of simple physically plausible models to represent complex biochemical processes. 

Secondly, the simplified model was developed using data relevant to normal physiological conditions. Most significantly, simplifying assumptions have been made about the physiological regime in which the model operates, and hence perturbations to species concentrations and specific enzymatic regulators outside of this regime are not captured in the simplified model. Full mechanistic exploration of mitochondrial dynamics under a broad range of perturbations and experimental data beyond those used to fit the simplified model should of course be pursued using the full bond graph representation of mitochondrial bioenergetics. 

Finally, because the physically-plausible model is energy based, any or all of the \Ce components can provide connections with energy-based models of other parts of the mitochondria system such as the TCA cycle, ATPase and ROS generation. In bond graph terminology, the \Ce components become \emph{ports} \citep{GawCra18} with which to connect to other bond graph components representing other aspects of mitochondrial biochemistry, for simulation and analysis of larger-scale models of mitochondrial and cellular bioenergetics. In our future work we intend to  exploit this feature of bond graphs to investigate in larger models of mitochondrial function the basis of ROS generation and damage.

\end{New}

\section*{Acknowledgements}
PJG would like to thank the Melbourne School of Engineering
for its support via a Professorial Fellowship.
This research was in part conducted and funded by the Australian
Research Council Centre of Excellence in Convergent Bio-Nano Science
and Technology (project number CE140100036).
The authors would like to thank an anonymous reviewer for pointing
out an error in an early version of the manuscript and suggesting a number of
points of clarification.

\end{document}